\documentclass[manuscript]{acmart}
\usepackage{comment}
\def\BibTeX{{\rm B\kern-.05em{\sc i\kern-.025em b}\kern-.08emT\kern-.1667em\lower.7ex\hbox{E}\kern-.125emX}}
    
\begin{document}








%
\title{SoK: Why Johnny Can't Fix PGP Standardization}

%
\author{Harry Halpin}
\email{harry.halpin@inria.fr}
\orcid{0003-2143-6965}
\affiliation{
  \institution{Inria}
  \streetaddress{2 Rue Simone Iff}
  \city{Paris}
  \state{France}
}


%

%
\begin{abstract}
 Pretty Good Privacy (PGP) has long been the primary IETF standard for encrypting email, but suffers from widespread usability and security problems that have limited its adoption. As time has marched on, the underlying cryptographic protocol has fallen out of date insofar as PGP is unauthenticated on a per message basis and compresses before encryption. There have been an increasing number of attacks on the increasingly outdated primitives and complex clients used by the PGP eco-system. However, attempts to update the OpenPGP standard have failed at the IETF except for adding modern cryptographic primitives. Outside of official standardization, Autocrypt is a ``bottom-up'' community attempt to fix PGP, but still falls victim to attacks on PGP involving authentication. The core reason for the inability to ``fix'' PGP is the lack of a simple AEAD interface which in turn requires a decentralized public key infrastructure to work with email. Yet even if standards like MLS replace PGP, the deployment of a decentralized PKI remains an open issue.
\end{abstract}

%
%


%
\keywords{email, PGP, standards, encryption}

%

%
\maketitle

\section{Introduction}

Although it has been over two decades since its inception as an open standard, the OpenPGP (Pretty Good Privacy) standard is still widely deployed despite issues with usability, key management, and numerous attacks. The most recent process of standardization of OpenPGP at the IETF (Internet Engineering Task Force) started in 2015 and ended in November 2017, but did not attempt to address these underlying issues, instead focusing on upgrading the core primitives to use modern cryptographic primitives -- a task which even the IETF standards process could not fully fix. After years of unsuccessful attempts to address PGP problems trying to make PGP more functional via new advanced server-side architecture~\cite{leap} and Working Group activities at the IETF, a number of OpenPGP client developers created a new community effort called ``Autocrypt'' to address the underlying usability and key management issues. This effort also introduces new attacks and does not address some of the underlying cryptographic problems in PGP, problems that have been addressed in more modern protocol designs like Signal or IETF Message Layer Security (MLS). After decades of work, why can't the OpenPGP standard be fixed? 

First, we start with the history of standardization of OpenPGP in Section~\ref{pgp:standards}. We consider the PGP protocol itself according to the modern understanding of cryptography in Section~\ref{pgp:protocol}, inspecting whether some original design choices still make sense in terms of the security and privacy properties a user would expect. Then we move to summarizing the numerous attacks on PGP that the design and incomplete standardization of PGP opens up in Section~\ref{pgp:attacks}. Note that there has been a large body of usability work on PGP, mostly with negative results and a large number of attacks based purely on usability problems, that we will only briefly overview as our primary focus in on standardization and protocol issues with PGP~\cite{johnny}. Lastly, we analyze the failed attempts to fix these issues via standardization in Section~\ref{pgp:failed}, including informal community-driven approaches like Autocrypt. Given the plethora of issues that exist with PGP, we analyze whether or not OpenPGP as a standard should continue at the IETF, given the underlying problem is a lack of a decentralized public key infrastructure in Section~\ref{conclusion}.




\section{Standardization of OpenPGP}
\label{pgp:standards}

Pretty Good Privacy (PGP) is a well-known cryptographic protocol for using asymmetric cryptography to provision confidentiality to the masses. At the very time when the U.S. government was attempting to restrict the spread of ``strong'' encryption in what was known as the first ``Crypto Wars,'' the release of PGP's open source code made it seemingly impossible to control the spread of tools for encryption~\cite{singh1999code}.  

The original scheme of PGP (``PGP 1.0'') was written by Philip Zimmerman, a peace activist, with its source code published on the Internet in 1991, first to a ISP called PeaceNet and then to Usenet. The goal was encrypting email, based on the plaintext Simple Mail Transfer Protocol (SMTP).  As PGP depended on RSA primitives for public key encryption and signatures, PGP attracted the attention of the company RSA Security Inc., which had patented the original RSA algorithms -- and even the United States government, which attempted to prosecute Zimmerman on breaking an encryption ban on ``strong'' cryptography by posting the source code internationally.

Turning PGP into a standard at the IETF both helped prevent a single company from preventing its further spread via patent claims by providing a version that did not use RSA (thus the standard was called ``OpenPGP'') and served as a strategic bulwark against further repression by weaving encryption into the open standards that defined the Internet itself.\footnote{Much of the history of PGP and its standardization has been summarized from various mailing lists on \emph{ietf.org}.}  Although Ron Rivest and others at MIT convinced RSA Security Inc. not to pursue any further patent actions and the United States government dropped their court case against Zimmerman in January 1996,  Phil Zimmerman submitted an informational (non-standard) RFC 1991 called ``PGP Message Exchange Formats''\footnote{\url{https://tools.ietf.org/html/rfc1991}} to the IETF in August 1996, describing various minor versions of ``PGP 2.0.'' Importantly, this document divided the actual encrypted message into separate variable length ``packets,'' where each packet contained a header specifying its type (``tag'') and a packet body. These packets may contain encrypted keys, encrypted messages, signatures, certificates, and so on. RFC 1991 featured not only RSA, but also the patented IDEA algorithm\footnote{\url{https://register.epo.org/application?number=EP91908542}} and MD5 for hash functions.\footnote{Although these may seem questionable choices today given various known attacks on IDEA and MD5, it should be remembered that cryptography was generally not standardized at the time with both IDEA and MD5 considered secure. Zimmerman replaced his own broken cipher Bass-o-matic in PGP 1.0 with IDEA after the release of PGP.}

In order to harmonize with email standards, OpenPGP was made compatible with Multi-purpose Internet Mail Extensions (MIME)\footnote{The IETF standard for sending messages with multiple components (such as a digital signature and a body as specified in IETF RFC 1521 \url{https://tools.ietf.org/html/rfc1521}} via IETF RFC 2015 ``MIME Security with Pretty Good Privacy.''\footnote{\url{https://tools.ietf.org/html/rfc2015}}  While OpenPGP-encrypted email was originally delivered simply as encrypted and signed plaintext in the OpenPGP Message format using standard SMTP, email best practices mandated some parts be delivered as multi-part MIME messages. In MIME messages, the content of the message is divided between different MIME parts, so that the encrypted body is delivered as a separate MIME part than the signature or attached keys.

The year after the original submission by Zimmerman, the original IETF OpenPGP Working Group opened in September 1997. Within a year the OpenPGP produced IETF RFC 2440 in 1998\footnote{\url{https://tools.ietf.org/html/rfc2440}} on ``the OpenPGP Message Format.''\footnote{Confusingly, this standard described PGP 5.0, because the rights to PGP 4.0 (and all even numbered versions of PGP) was bought by Viacrypt (which became PGP Incorporated), PGP 3.0 evolved PGP 5.0 as Zimmerman held the rights to odd numbered PGP versions.}   This document also described certificates, identities, and in more detail the operations necessary for signing, encryption, and decryption of data with OpenPGP. In particular, it expanded the ciphersuite choice by making mandatory the patent-free El-Gamal for asymmetric encryption, SHA-1 for hash functions, DSA for signatures, and 3DES for symmetric encryption. It also ended up recommending CAST-128, with an infamous 128 bit key and the weak S2K (String-to-Key) KDF for converting plaintext passwords to keys. MIME usage was also updated in 2001 as RFC 3156 ``MIME Security with Open PGP.''\footnote{\url{https://tools.ietf.org/html/rfc3156}} The OpenPGP Working Group was then closed.

Due to advances in cryptography and attacks  detailed in Section~\ref{pgp:attacks}, the IETF reopened the OpenPGP Working Group in May 2003. Four years later RFC 2440 was updated by the Working Group as IETF RFC 4880 in 2007.\footnote{\url{https://tools.ietf.org/html/rfc4880}}  The primary change was the addition of PKCS1-v1\_5 encoding and the optional use of a modified version of AES-CFB mode for symmetric encryption with the addition of Modification Detection Codes (MDCs) to provide a limited form of integrity to symmetric encryption. In brief, an MDC attempts to alert the recipient if their message has been corrupted by attaching an encrypted SHA-1 hash of the plaintext as the last packet in a message. Also, MD5 was deprecated and IDEA made optional.  

The OpenPGP Working Group was again closed in 2008, but the mailing list continued to be somewhat active, with PGP able to work with most major operating systems, including eventually  Microsoft Outlook, Apple Mail, Mozilla Thunderbird, and others. The  majority of PGP implementations eventually ended up depending on the GNU Privacy Guard (GPG) library.\footnote{https://gnupg.org/} Although the cryptographic primitive choices seems questionable today and many options were given, it should be remembered that OpenPGP was built during a period when cryptographic standardization was just starting, and so simply including every plausible algorithm seemed to make sense given the security level of many algorithms was unknown. Also, cryptographic protocol design was also in its infancy, and so the PGP protocol itself is historic as an initial, if ad-hoc, ``real world'' open source cryptographic deployment.

\section{PGP Protocol Design}
\label{pgp:protocol}

\subsection{Basic PGP Protocol}

The core design of PGP is more than two decades old, and despite outstanding usability and key management issues, PGP is still used for encryption and signatures, although it does not have clear formal security properties and assumptions.

The PGP protocol can encrypt and sign a plaintext message $m$. This requires the sender having a private asymmetric key $sk$ and the recipient public key $pk$.

\begin{enumerate}
\item A signature $\sigma_m = SIGN(sk, m)$ is created with the key $sk$ .
\item By default, $m$ is compressed to $m'$. The compressed message $m'$ is concatenated with the signature of its uncompressed version, $\sigma_m$.
\item  A fresh new symmetric key $k$ is generated and encrypts the message such that the ciphertext $c_m = E(k, m' | \sigma_m)$.
\item This symmetric key $k$ is encrypted with $pk$ so $c_k  = E(pk, k)$.
\item The resulting encrypted message $c$ is the concatenation of the encrypted symmetric key and ciphertext, i.e. $c  = c_k | c_m$.
\end{enumerate}


The creation of a fresh random symmetric key $k$ in order to encrypt the actual plaintext message later became best practice. By the generation of the randomized symmetric key per message, PGP achieves probabilistic encryption for the message itself; each encryption of the same plaintext results in an unlinkable new ciphertext even if a deterministic encryption scheme is used. Speed is also achieved by depending primarily on symmetric encryption. The original message that is symmetrically encrypted is unbounded in length, while the symmetric key $k$ always the same size and so can be asymmetrically encrypted relatively quickly. Lastly, PGP allows group messages as used in email efficiently, as the same encrypted message $c_m$ can simply be sent with different $c_k$ for the new public key $pk$ of each recipient, so the plaintext is not encrypted multiple times.

\subsection{Design Issues in OpenPGP}
\label{pgp:design}

Nonetheless, PGP does make some design choices that simply do not make sense in terms of modern cryptographic protocol design, regardless of the primitives used or efficiency gains. The highlights of the fundamental problems are given below.

\subsection{Message Compression}

Hiding the length of a plaintext message is not typically a security property but a privacy property, and so many protocols do not defend against attacks on the length of the ciphertext, including PGP. Furthermore, when standards like OpenPGP and TLS 1.2  were created, due to bandwidth considerations the cleartext was often compressed. What was not known at the time was that message compression can lead to attacks on encrypted data as information about plaintext entropy can gained from the information leaked by the compressed length~\cite{kelsey}, and so compression is no longer used in standards like TLS. Compression attacks work by virtue of repeated transmission of (nearly) the same message with only slight modifications by the attacker, and having those messages cause a response to be returned that includes the encrypted value that the attacker wants to find in a part of the message that is not controlled by the attacker. In practice rate-limiting can prevent this attack (unlike in TLS) and the kinds of secrets it would likely be able to find are limited -- the attack does not lead to a decryption of the content of the message, but could allow an attacker to guess authentication tokens or secret links like ``unsubscribe'' in parts of messages. When OpenPGP-level compression is explicitly disabled, applications for which the plaintext length is sensitive should implement some application-specific padding scheme in order to disguise the length of the plaintext. These sorts of techniques are already used by packet formats such as the Sphinx format for mix networking so that all messages have uniform length~\cite{danezis2009sphinx}. This can be done in a more lightweight manner for OpenPGP by simply adding (even randomized) padding to the uncompressed message. Currently, OpenPGP also does not encrypt the sender and other information not needed for message routing, and proposals like PURBs manage to hide all fields not needed for routing and the length of messages~\cite{nikitin2019reducing}. 

\subsection{Sign-then-encrypt}

The second issue is that OpenPGP itself functions as a library that allows any use of encryption and signing as API calls. For example, many users do not sign any messages at all as they believe these signatures may leak metadata about them. By default, strangely OpenPGP uses a sign-then-encrypt scheme so the recipient of the message is not authenticated, only the sender. Therefore even when a message is signed, sign-then-encrypt allows surreptitious forwarding attacks~\cite{davis}, where the legitimate recipient of a message signed by the sender can re-encrypt the message to an unlimited number of arbitrary recipients of their choice -- while the original signature of the sender remains valid! This likely violates a user's expectation both about how messages and signatures work. These ``spoofing'' attacks that alter messages en route between email providers both easily possible although surprisingly not widespread outside of spam, and could target OpenPGP-encrypted messages~\cite{hu2018end}. To concretely illustrate, imagine that Bob sends the signed message ``I love you'' to Alice using OpenPGP encryption. Alice can decrypt the message, re-encrypt it and send it to Carol. Carol may obviously misinterpret the message as a statement that Bob loves her, not Alice - and the cryptographic assurances of the digital signature are violated. Alice can even make it seem that the message came directly from Bob to Carol. While not devastating, this attack could lead to very surprising results, particularly in terms of encrypted group messaging where it would allow anyone to arbitrarily expand the group while the ``appearance'' of encryption and digital signatures would remain the same to the original members.  This problem of a lack of clear signing authority over a given message can also lead to many attacks on the difficulty of determining the authority behind the signer or which particular message has a particular signer in multi-part and nested messages, a situation exacerbated by PGP's unlimited nesting of signed messages~\cite{fired}.

\subsection{Unauthenticated Headers}

Not only can signatures be stolen, but all SMTP headers are unauthenticated by OpenPGP. Indeed, the surreptitious forwarding attack is made even easier due to this underlying issue~\cite{muller2019re}. For example, the \emph{From:} header specifies the sender and this identity should match the identity of the OpenPGP identifier of the sender in the signature, yet this is not done. In practice, DomainKeys Identified Mail (DKIM) enabled mail providers and even mail agents agents can already match the identity of DKIM email signature with the \emph{From:} field value, and display a warning if a mismatch is found from the originating domain.\footnote{\url{https://tools.ietf.org/html/rfc6376}} Unfortunately, not all MUAs perform this check~\cite{liao}, leaving some users exposed to impersonation attacks. Regardless, it is only the sender's domain, not the user, that is checked by DKIM. So, an impersonation attack that replaces  \emph{bob@gmail.com} with \emph{eve@gmail.com} would work even with DKIM-certified mail. The \emph{Subject:} field is also sensitive and can be tampered with. A few encrypted email agents have began removing subjects by default and replacing them with strings such as ``...'' and putting the header in the body using a technique called ``memory hole.''\footnote{\url{https://0xacab.org/leap/pymemoryhole}} Lastly, altering the \emph{Reply-To:} field could allow an adversary to redirect the response to an inbox of their choosing, and email users may be unaware that emails do not return to the address in the \emph{From:} field automatically. Historically this field can be used to respond to mailing lists and current encrypted mailing list software like Schleuder\footnote{\url{https://schleuder.org/}} use a trusted third-party server to manage a single long-term RSA keypair for the mailing list, and this field could be tampered with in order to prevent messages from returning to the correct mailing list even if encryption is used.

These are only the most abstract design issues with PGP. There are even more problems with PGP on a less fundamental level. For example, the use of MDC codes being attached at the end of a message allows them to be easily stripped by an attacker. Worse, it also prevents large files from being streamed with PGP, as the MDC is checked only at the end of the file. How these design flaws play out in terms of concrete attacks is shown in the next section.

\section{Attacks on PGP}
\label{pgp:attacks}

\subsection{Protocol Attacks}

Even without directly attacking the known broken cryptographic primitives supported by OpenPGP due to ``backwards compatibility,'' there have been a number of attacks on PGP's cryptographic foundations. For example, one of the earliest academically published attacks was to use a chosen ciphertext attack, where an adversary would send an encrypted message that could not be decrypted and then the user would respond, quoting the ciphertext sent by the adversary in the response~\cite{katz2000chosen}.  This would allow the adversary to know if their ciphertext could be decrypted or not, which would eventually lead to the breaking of the encryption itself, which was shown to be the case in early versions of OpenPGP-compliant software~\cite{jallad2002}. Another attack was done on the PGP's changes to AES-CFB mode (i.e. ``PGP-CFB'' mode) that were made partially in response to the previous attack~\cite{jallad2002}. PGP could leak significant bits of the plaintext via an attack that takes advantage of a idiosyncratic ``quick check'' for session key correctness added to PGP-CFB mode, although the attack works only if timing attacks can be exploited quickly, so ``quick check'' bytes are simply not checked in newer implementations~\cite{mister2005attack}. Although RSA PKCS \#1 v1.5 padding was mandated for the RSA keys in OpenPGP, the more recent IETF RFC also mandated that the errors in key processing return the same error the clients as decryption errors, so surprisingly it appears all modern PGP clients are not vulnerable to Bleichenbacher's attack~\cite{fired}.

\subsection{Client Attacks}
 
The usage of complex MIME types has been the source of more recent attacks on the modern OpenPGP standard. The problem is email clients handle the PGP MIME types differently, with some clients allowing the handling of rich HTML content with Javascript and CSS for messages. As HTML content with Javasript allows the insertion of arbitrary remote content without proper security boundaries (which many email clients do not enforce), the ``eFail'' attack allowed PGP messages to have their plaintext content captured and sent to an adversary operating a remote server via wrapping the ciphertext in a MIME part controlled by an adversary and also taking advantage of problems caused by compression~\cite{efail}. This attack took advantage of a failed MDC code to reveal the plaintext of a message. This was not due to a failure in the standard, clients implementing the standard correctly, namely the suggestion of RFC 4880 that in the case of a failed MDC that an ``implementation[s] MAY allow the user access to the erroneous data'' with a warning.

The lack of authentication in the PGP protocol leads to real-world surreptitious forwarding attacks that take advantage of the poor usability and conflicting client behavior in OpenPGP-compatible clients. In general, these attacks cause a user to mistakenly believe a PGP message that has a detached signature (of an ``Encrypt-then-Sign'') in a separate MIME part to appear to be a valid signature for in an entirely different MIME part in some clients~\cite{fired}. For example, an adversary can take a valid encrypted message with a signature attached via MIME, strip the original signature, and simply insert their own signature MIME part, forwarding the entire PGP/MIME message to their victim whose client UI mistakenly shows the attacker's signature as valid for the original message. These attacks are not confined to PGP/MIME processing, but also can happen with inline PGP signatures that do not use MIME types but embed the signature directly in the text of an email without a MIME separation of content due to the earlier issues on ``sign-then-encrypt'' combined with unlimited signature nesting. Until 2007 arbitrary unsigned data could be injected into to a message with an inline signature and the entire message would validate as signed in PGP~\cite{fired}. In inline messages, content at the beginning of a message could be changed without invalidating the signature.~\footnote{\url{https://dkg.fifthhorseman.net/notes/inline-pgp-harmful/}} 

\subsection{Key Management Attacks}

The lack of well-specified key management in OpenPGP has also led to numerous attacks on the common ``folk practice'' used in OpenPGP for key management. Keys in PGP are usually not transmitted as the raw bytes of the key itself, but via a PGP certificate as specified that matches each key to a User ID, a string that may contain an email address. The PGP key certificate is for a signing key that then self-signs the entire certificate, verifying the binding between the User ID and one or more keys. Also for each key there is a key ID (a truncated number of bits of the public key), PGP version number, a revocation status, and possibly expiration date as well as a host of optional fields such as an (often unused) field for the preferred key server. Each PGP certificate for a key may contain multiple sub-keys, which are full-fledged keys in their own right used for either encryption or signing, where the signature of a key also applies to a subkey of that key. Soon after PGP itself was launched by Zimmerman, a key server\footnote{\url{https://pgp.mit.edu}} was launched at MIT by Brian LaMacchia  to allow the discovery of keys via lookup via User IDs and Key IDs.

There were three critical flaws with key servers, whose operations were never standardized. First, there is no requirement that Key IDs and User IDs be unique, so attacks such as the ``Evil 32'' attack showed that practical collusion were possible with 32 bit Key IDs in 4 seconds.\footnote{\url{https://evil32.com/}} More dangerously, there was no authentication used by key servers, so anyone could upload a key to any key server claiming to be anyone. HTTP Key Server (HKS) allows the maintenance of keys,\footnote{\url{https://tools.ietf.org/html/draft-shaw-openpgp-hkp-00}} and keys servers can share keys using Synchronized Key Servers (SKS).\footnote{\url{https://sks-keyservers.net/}} As key servers allowed the updating of keys, anyone could update a key belonging to anyone else. In 2019, two OpenPGP designers were attacked by a ``certificate flooding'' attack where an adversary maliciously added a large amount of keys to the key certificates of a user on one or more key servers, effectively rendering the certificates incapable of being processed.\footnote{\url{https://access.redhat.com/articles/4264021}} Lastly, key verification is in theory done via a ``Web of Trust'' with multiple trust levels (unknown, none, marginal, full, and ultimate) that are interpreted by users differently, and the ``Web of Trust'' forms a ``small world'' network that essentially leads to the same kinds of trust as a centralized PKI.\footnote{\url{https://pgp.cs.uu.nl/plot/}} User key material on key servers, including their social graph of ``trust,'' is completely public and so reveals the personal data of everyone's web of trust, which seems against any privacy properties.

\section{Autocrypt: Fixing PGP?}
\label{pgp:failed}

When there are well-known attacks on an IETF standard, such as the numerous attacks on TLS 1.2, the standard is usually upgraded by a rechartered IETF Working Group. However, many in the community and IETF did not think PGP needed upgrading as the majority of attacks outlined in Section~\ref{pgp:attacks} are not cryptographic attacks on PGP itself, but primarily due to OpenPGP's interaction with a wider and mostly under-specified infrastructure such as email clients and key servers. 

An attempt to provide a list of sensible changes to address problems with PGP, such as using AES-GCM or ChaCha20 with Poly1305 for symmetric encryption, as well as improve the metadata properties of the OpenPGP format, were given the ``Modernizing the OpenPGP Message Format'' draft.\footnote{\url{https://tools.ietf.org/html/draft-ford-openpgp-format-00}}  The OpenPGP Working Group was re-opened for a third time in 2015, and the IETF  restricted its charter to updating the primitives to use modern elliptic curve cryptography and AEAD encryption for the underlying encrypted data that were at the time protected by the SHA1-based MDC.\footnote{Earlier in 2012, NIST curve compatibility as given by IETF RFC 6637: \url{https://tools.ietf.org/html/rfc6637}} Although there were drafts to add EdDSA for signatures\footnote{\url{https://tools.ietf.org/html/draft-koch-eddsa-for-openpgp-04}} and a revised RFC4880bis that used Curve 25519 as well as AEAD ciphersuites,\footnote{\url{https://tools.ietf.org/html/draft-ietf-openpgp-rfc4880bis-07}} the OpenPGP Working Group at the IETF was closed in November due to lack of community involvement and review of the proposed standard. The mailing list and updates to RFC4880bis are still active, but this community has no mandate by the IETF to create an official standard. The previous co-chairs of the Working Group have suggested if energy could be found to finalize RFC 4880bis that the OpenPGP Working Group could be re-chartered.\footnote{Personal communication, Barry Leiba, former IETF OpenPGP Working Group chair.}

Attacks on key servers and PGP certificates would require updating the PGP certificate part of RFC 4880bis as well as specifying how key servers work. Non-standardized guidance has been presented to prevent these attacks that pre-dated the attacks themselves, but the document runs over fifty pages, pointing to the seemingly impossible security and privacy issues with existing PGP key servers.\footnote{\url{https://tools.ietf.org/html/draft-dkg-openpgp-abuse-resistant-keystore-03}} The OpenPGP Working Group drafted a proposed key server draft at the IETF which verified keys uploaded via TLS by sending an encrypted nonce from the key server to the user's email and receiving the signed nonce in a reply email but there have been no implementations.\footnote{\url{https://tools.ietf.org/html/draft-koch-openpgp-webkey-service-08}} A number of ambitious projects were initiated to fix the flaws in key discovery and management in PGP like Pretty Easy Privacy,\footnote{\url{https://tools.ietf.org/id/draft-marques-pep-email-02.html}} and the Leap Encryption Access Project (LEAP) to automate key discovery by relying on a federation of email servers~\cite{leap}. These approaches were not deployed as existing email servers did not wish to upgrade their existing server-side infrastructure to handle key management.

The goal of new \emph{Autocrypt} community effort is to allow keys to be transmitted not via OpenPGP keyservers, but via the emails themselves.\footnote{\url{https://autocrypt.org/}} Emails should default to encrypted where possible, and so it is better to be confidential ``most of the time'' rather than not. This requires ``Trust on First Use'' (TOFU) messages that, while defaulting to cleartext, attempt to upgrade to encryption as soon as possible in the flow of the protocol. Autocrypt does not alter the cryptographic constructions or primitives, but instead adds Autocrypt-specific mail headers in email clients that are backwards compatible with all OpenPGP clients and do not cause email users who do not use OpenPGP to become aware of encryption. Email users who do not use OpenPGP are naturally confused by receiving an encrypted message or key material as an attachment using PGP/MIME, and users of encrypted email can receive messages they cannot open due to expired or incorrect keys being used. Autocrypt solves this problem as headers that are not understood by an existing mail client will just be dropped, and so not displayed to the user. Details are given in the Autocrypt specification 1.1.\footnote{\url{https://autocrypt.org/autocrypt-spec-1.1.0.pdf}} 

The Autocrypt community is not an official standard, but a ``bottom-up'' community-driven process led by email client developers. They released the first version of the ``Autocrypt Level 1'' specification in December 2017 (after the IETF closed the OpenPGP Working Group), which was updated to version 1.1 at the end of 2019. The primary change between version 1.0 and version 1.1 was the movement of the default from large RSA keys for signing and encryption to Ed25519 and Curve 25519 keys. Currently, a number of different OpenPGP-compliant clients also support Autocrypt, in particular Enigmail, K-9 mail, mutt, and eight other OpenPGP email clients. The core problem with Autocrypt is that it does not address the underlying problems with lack of authentication in PGP, and so suffers from the same attacks on SMTP message spoofing due to unauthenticated headers. In particular, an attacker can spoof both the initial Autocrypt headers by creating an impersonated email header and force an Autocrypt-enabled client to insert the attacker's key in their keystore for an arbitrary contact. Due to Autocrypt's key updating mechanism, an attacker can simply replay this attack via faking unauthenticated headers to replace a communication peer's key with an attacker's key or gossip incorrect new keys for users. Given the promotion of Autocrypt usage via security messaging using PGP with Delta.chat in places like Russia and Ukraine, these problems may become serious.\footnote{\url{https://delta.chat}} The danger is that users believe their messages are confidential and even authenticated, when they are indeed not~\cite{halpin}. Although the alternatives like Signal are centralized, they are likely more secure and inline with user expectations. Opportunistic encryption does simply not provide the certainty needed by many users about their communication.

\section{Conclusion: Can PGP be fixed?}
\label{conclusion}

The question facing users, developers, and security researchers is whether PGP should be considered harmful and thus deprecated from common usage and future standardization. As we have shown, many of the problems with PGP come from its lack of an authenticated encryption with additional data construction (AEAD)~\cite{aead}. A number of recommendations have been made on the level of the OpenPGP itself:
\begin{itemize}
\item Removing compression, as it is no longer needed due to the increased capacity of the internet and the only reason for keeping it, a chosen ciphertext attack, is prevented by the use of an AEAD symmetric ciphers like ChaCha20-Poly1305 and even MDCs.
\item Moving to modern elliptic curve cryptography for signing and public-key encryption and AEAD for symmetric encryption, as specified in the defunct 2015 OpenPGP charter. 
\item As MDCs are redundant if the message is signed or an AEAD ciper is used, as well as being hard-coded to be based on SHA1 and even lead to attacks~\cite{mister2005attack}, MDCs should be removed when a modern AEAD scheme is used for symmetric encryption.
\item Use AEAD not only on the OpenPGP message format, but email itself, with routing specific headers like \emph{From:, To:}, and \emph{Reply-To:} given integrity.
\end{itemize}

Yet as pointed out, the protocol itself is so flawed in terms of design (such as lacking forward secrecy~\cite{otr}\footnote{Although there have been attempts to add forward secrecy to OpenPGP such as \url{https://tools.ietf.org/html/draft-brown-pgp-pfs-03}}), it seems impossible to fix without breaking backwards compatibility and compatibility with email and MIME (in order to support AEAD natively). One option is simply to abandon fixing PGP and start from a cleaner design that natively supports modern cryptography and a per-message AEAD construction. In 2018, the IETF convened a new Working Group, Message Layer Security (MLS) to design an IETF standardized AEAD group messaging protocol that could scale to large groups, a problem the unstandardized Signal Protocol is unable to tackle due to its design.\footnote{\url{https://datatracker.ietf.org/wg/mls/about/}} Would it make sense to drop fixing OpenPGP for messaging entirely and concentrate only on MLS and usability in encrypted messaging? Perhaps, but let's not forget the larger picture:  Technically, decentralized public key infrastructure is an open research problem, both in terms of cryptography and usability. The usability of keys as identifiers is \emph{the} root usability problem of PGP, and without a breakthrough, new standards will not fix this unless keys are managed on behalf of the user -- which goes against the values of decentralization and user control. 

Simply deprecating OpenPGP without further work misses the real reason why there are still many supporters of OpenPGP for encryption in the developer community: OpenPGP is viewed as open and decentralized as opposed to current fragmented secure messaging applications. Decentralized key management would be needed for authentication in a decentralized environment, and a lack of this infrastructure explains the lack of usage of an AEAD construction in OpenPGP.  There has been considerable work on privacy-preserving and authenticated centralized key-servers for generic key material, such as CONIKS~\cite{melara2015coniks} and its implementation in Key Transparency.~\footnote{\url{https://github.com/google/keytransparency}} Although Key Transparency was originally going to be used for OpenPGP-compatible email between Google and Yahoo, the project was discontinued. Inspired by decentralized blockchain technology, ClaimChain presents a decentralized, gossip-based alternative to centralized key management without key servers with privacy~\cite{claimchain}. Autocrypt has already supported the sending of communication peer keys via gossip protocols.\footnote{\url{https://countermitm.readthedocs.io/en/latest/}} Decentralization via gossip has its drawbacks; the approach using a real-world data-set and could reach consensus only on  66\% of users in terms of key discovery~\cite{claimchain}. Further work on automated key discovery and access control based on decentralization, such as the nearly forgotten Simple Public Key Infrastructure standards\footnote{\url{https://tools.ietf.org/wg/spki/}} based on SDSI~\cite{sdsi}, could create more robust and privacy-preserving decentralized key management solutions. The real problem standards bodies can't fix PGP is not due to the legacy ciphersuites or even protocol details, but the hard problem of key decentralized management. 

\begin{acks}
Harry Halpin was funded in part by the European Commission H2020 through the NEXTLEAP Project (Grant No. 6882). He would like to thank Alfredo Pironti for his talk to the NEXTLEAP project members at IMDEA in March 2017 on PGP and then sharing his draft article ``Reconciling actual and expected security in OpenPGP,'' which inspired Section~\ref{pgp:protocol}, and looks forward for Pironti developing his work further. He would also like to thank Holger Krekel,  Barry Leiba, and Daniel Kahn Gilmor for discussions on PGP and Autocrypt, although they may not agree with the conclusions given in this paper. 
\end{acks}

\bibliographystyle{plain}
\bibliography{main}

\begin{thebibliography}{10}

\bibitem{otr}
Nikita Borisov, Ian Goldberg, and Eric Brewer.
\newblock Off-the-record communication, or, why not to use {PGP}.
\newblock In {\em Proceedings of the 2004 ACM workshop on Privacy in the
  electronic society}, pages 77--84. ACM, 2004.

\bibitem{danezis2009sphinx}
George Danezis and Ian Goldberg.
\newblock Sphinx: A compact and provably secure mix format.
\newblock In {\em 2009 30th IEEE Symposium on Security and Privacy}, pages
  269--282. IEEE, 2009.

\bibitem{davis}
Don Davis.
\newblock Defective sign \& encrypt in {S/MIME}, {PKCS}\# 7, {MOSS}, {PEM},
  {PGP}, and {XML}.
\newblock In {\em USENIX Annual Technical Conference, General Track}, pages
  65--78, 2001.

\bibitem{halpin}
Harry Halpin, Ksenia Ermoshina, and Francesca Musiani.
\newblock Co-ordinating developers and high-risk users of privacy-enhanced
  secure messaging protocols.
\newblock In {\em International Conference on Research in Security
  Standardisation}, pages 56--75. Springer, 2018.

\bibitem{hu2018end}
Hang Hu and Gang Wang.
\newblock End-to-end measurements of email spoofing attacks.
\newblock In {\em 27th USENIX Security Symposium}, pages 1095--1112, 2018.

\bibitem{jallad2002}
Kahil Jallad, Jonathan Katz, and Bruce Schneier.
\newblock Implementation of chosen-ciphertext attacks against {PGP} and
  {G}nu{PG}.
\newblock In {\em International Conference on Information Security}, pages
  90--101. Springer, 2002.

\bibitem{katz2000chosen}
Jonathan Katz and Bruce Schneier.
\newblock A chosen ciphertext attack against several e-mail encryption
  protocols.
\newblock In {\em USENIX Security Symposium}, 2000.

\bibitem{kelsey}
John Kelsey.
\newblock Compression and information leakage of plaintext.
\newblock In {\em International Workshop on Fast Software Encryption}, pages
  263--276. Springer, 2002.

\bibitem{claimchain}
Bogdan Kulynych, Wouter Lueks, Marios Isaakidis, George Danezis, and Carmela
  Troncoso.
\newblock Claimchain: improving the security and privacy of in-band key
  distribution for messaging.
\newblock In {\em Proceedings of the 2018 Workshop on Privacy in the Electronic
  Society}, pages 86--103. ACM, 2018.

\bibitem{liao}
Lijun Liao and J{\"o}rg Schwenk.
\newblock End-to-end header protection in signed {S/MIME}.
\newblock In {\em OTM Confederated International Conferences" On the Move to
  Meaningful Internet Systems"}, pages 1646--1658. Springer, 2007.

\bibitem{melara2015coniks}
Marcela~S Melara, Aaron Blankstein, Joseph Bonneau, Edward~W Felten, and
  Michael~J Freedman.
\newblock {CONIKS}: Bringing key transparency to end users.
\newblock In {\em 24th USENIX Security Symposium}, pages 383--398, 2015.

\bibitem{mister2005attack}
Serge Mister and Robert Zuccherato.
\newblock An attack on {CFB} mode encryption as used by {O}pen{PGP}.
\newblock In {\em International Workshop on Selected Areas in Cryptography},
  pages 82--94. Springer, 2005.

\bibitem{fired}
Jens M{\"u}ller, Marcus Brinkmann, Damian Poddebniak, Hanno B{\"o}ck, Sebastian
  Schinzel, Juraj Somorovsky, and J{\"o}rg Schwenk.
\newblock Johnny, you are fired! -- spoofing {O}pen{PGP} and {S/MIME}
  signatures in emails.
\newblock In {\em 28th USENIX Security Symposium}, 2019.

\bibitem{muller2019re}
Jens M{\"u}ller, Marcus Brinkmann, Damian Poddebniak, Sebastian Schinzel, and
  J{\"o}rg Schwenk.
\newblock Re: {W}hat's up {J}ohnny?
\newblock In {\em International Conference on Applied Cryptography and Network
  Security}, pages 24--42. Springer, 2019.

\bibitem{nikitin2019reducing}
Kirill Nikitin, Ludovic Barman, Wouter Lueks, Matthew Underwood, Jean-Pierre
  Hubaux, and Bryan Ford.
\newblock Reducing metadata leakage from encrypted files and communication with
  {PURBS}.
\newblock {\em Proceedings on Privacy Enhancing Technologies}, 2019(4):6--33,
  2019.

\bibitem{efail}
Damian Poddebniak, Christian Dresen, Jens M{\"u}ller, Fabian Ising, Sebastian
  Schinzel, Simon Friedberger, Juraj Somorovsky, and J{\"o}rg Schwenk.
\newblock Efail: Breaking {S/MIME} and {O}pen{PGP}s email encryption using
  exfiltration channels.
\newblock In {\em 27th USENIX Security Symposium}, pages 549--566, 2018.

\bibitem{sdsi}
Ronald Rivest and Butler Lampson.
\newblock {SDSI}-a simple distributed security infrastructure.
\newblock Technical report, MIT, 1996.

\bibitem{aead}
Phillip Rogaway.
\newblock Authenticated-encryption with associated-data.
\newblock In {\em Proceedings of the 9th ACM conference on Computer and
  communications security}, pages 98--107. ACM, 2002.

\bibitem{singh1999code}
Simon Singh.
\newblock {\em The code book}.
\newblock Doubleday New York, 1999.

\bibitem{leap}
Elijah Sparrow, Harry Halpin, Kali Kaneko, and Ruben Pollan.
\newblock {LEAP}: A next-generation client {VPN} and encrypted email provider.
\newblock In {\em International Conference on Cryptology and Network Security},
  pages 176--191. Springer, 2016.

\bibitem{johnny}
Alma Whitten and J~Doug Tygar.
\newblock Why {J}ohnny {C}an't {E}ncrypt: {A} usability evaluation of {PGP}
  5.0.
\newblock In {\em {USENIX} Security Symposium}, volume 348, pages 169--184,
  1999.

\end{thebibliography}

\end{document}